# The Challenges with Internet of Things for Business

Ievgeniia Kuzminykh[1], Bogdan Ghita[2], and Jose M. Such[1]

***Abstract*— Many companies consider IoT as a central element for increasing competitiveness. Despite the growing number of cyberattacks on IoT devices and the importance of IoT security, no study has yet primarily focused on the impact of IoT security measures on the security challenges. This paper presents a review of the current state of security of IoT in companies that produce IoT products and have begun a transformation towards the digitalization of their products and the associated production processes. The analysis of challenges in IoT security was conducted based on the review of resources and reports on IoT security, while mapping the relevant solutions/measures for strengthening security to the existing challenges. This mapping assists stakeholders in understanding the IoT security initiatives regarding their business needs and issues. Based on the analysis, we conclude that almost all companies have an understanding of basic security measures as encryption, but do not understand threat surface and not aware of advanced methods of protecting data and devices. The analysis shows that most companies do not have internal experts in IoT security and prefer to outsource security operations to security providers.**

## I. Introduction

GLOBAL market and society are currently undergoing the process of digitalization of the objects. Internet of things, big data, blockchain are all evidence of a global trend of moving valuables and activities from the physical world to the digital world that drive growth of the business and raise the competitiveness. IoT came to simplify and optimize business processes, improve society lives, allow people to control connected products, save money and time, while maintaining our security and privacy. Are companies ready for the secure transmission, processing, and storage of data collected by IoT services which are increasingly becoming part of their products and processes?

According to Cisco Annual Internet report we will have 29.3 billion networked devices by 2023 including smart TV, smartphones and M2M applications, such as smart meters, video surveillance, healthcare monitoring, transportation, and package or asset tracking [1].

The report Worldwide Global Data Sphere IoT Device and Data Forecast, 2019-2023, provides a forecast of 41.6 billion connected IoT devices, or “things”, generating 79.4 zettabytes (ZB) of data in 2025 [2].

But the level of security of new online technologies, including IoT, remains quite low. According to Gartner report in 2018 [3] most of the companies considered IoT security not as part of the business strategy but as line-of-business unit. Therefore, the poor “security by design”, and little control over the technology within connected devices were the consequences of the strategy and led to the growing number of cyberattacks on the IoT. In the period from 2015 to 2018 about 20% of the organizations were exposed to the attacks on IoT system, as reported by Gartner in their survey.

The number of cyberattacks on IoT devices is growing rapidly, as more and more customers, companies, municipal services start to use “smart” devices, such as routers, DVR cameras, smart traffic lights, asset trackers, smart meters, connecting to the Internet but not everyone is concerned about security [4]. By themselves, these devices may not be of interest to the cybercriminals. However, hackers crack them to use as robots to create botnets - networks of infected smart devices to conduct DDoS attacks - or as a proxy server for other types of malicious actions. Hackers simply need to discover the place where devices are connected not properly to be able to get into the system. And often, nine times out of ten they are successful. Most owners of hacked devices do not even suspect how their IoT devices are used. Cybercriminals see more and more financial opportunities to use such devices.

Regardless the number of attacks on IoT the Gartner report predicted that even in 2020 the security of the Internet of things would not be a priority for business [3]. In addition, the implementation of best security practices and tools in IoT planning would be ignored. Due to these two constraints, the companies can lose their reputation.

In this paper, we will take a look on the current state of the IoT security of the companies by analyzing the resources and available documentation on security in IoT. The purpose of the study is to identify and make analysis of the challenges that enterprises are faced when they plan and deploy IoT security at their products and processes. Moreover, several solutions to reduce risks related to IoT security have been analyzed as well, and been mapped against identified issues.

Despite the growing importance of IoT, no study has yet primarily focused on the impact of IoT security on the business strategy or business models. For example, Z. Bi et al in [5] investigated the impact of IoT on manufacturing and enterprises, K. Wnuk and B.Teja in [6] analyzed the impact of IoT on software business and requirements engineering, H. C. Y. Chan in [7] made analysis of value chain elements and stakeholders for IoT business model and validated proposed business models through the case studies of some companies. But none of these works had considered IoT security factor when developing or analyzing the impact of IoT on the business strategy or business model of enterprise. Our study is intended to fill this gap.

[1] King’s College London, UK (e-mail: ievgeniia.kuzminykh@kcl.ac.uk).
[2] University of Plymouth, UK

The organization of the paper is as follows: Section 2 describes threat surface related to the IoT devices and common vulnerabilities that lead to the attacks. Section 3 presents most important challenges raised with IoT security for businesses. In Section 4, the possible solutions to strengthen the IoT security are described, and analyzed in Section 5. Finally, Section 6 presents the conclusions.

## II. IoT Security Threat Landscape

Seventy six percent (76%) of small and large businesses now globally are adopting innovative digital cutting-age technologies like AI, IoT, cloud computing and automation to be more competitive, efficient and fast-moving. Some companies more focused on one or few technologies and less interested in another, statistic shows that the most popular is cloud computing (90%), and vast number of organization (76%) are considering to apply or already using IoT in their businesses [8]. But all these technologies bring additional cyber risk to sensitive data, connection, products and processes, using new technologies expands attack surface. This development makes companies more vulnerable to cyberattacks because of growing number of entry points [9]. Not properly implemented IoT security can lead to the data leakage, compromising of customers and business processes and loss of reputation for company. Moreover, IoT system in most cases is used for delivering malware or implementing more sophisticated attacks on the companies' or user's system, hackers can use the IoT network to launch attacks on regular computer networks in homes, businesses, or even smart cities.

Researchers at Check Point showed that networks can be hacked with a light bulb by testing market-leading Philips Hue smart lamps and controllers. They discovered vulnerabilities (CVE-2020-6007) that allowed them to penetrate the network using a remote exploit in the low-power ZigBee protocol which is used to control a large number of IoT devices. This vulnerability could allow the hacker to deliver ransomware or other malicious programs to office and home network [10].

The new version of the Mirai botnet at the end of 2019 targeted Zyxel devices [11] which had used factory-set default usernames and passwords and easy-to-guess password combinations. The cybercriminal posted open access lists of Telnet credentials with IP address, logins and passwords for more than 515 thousand servers, home routers and IoT devices. Such lists are often called bot lists as they are often used by IoT botnet operators. First, criminals scan the Internet making a list of bots, and then use it to connect to devices and infect them with malware.

At the beginning of 2020 researchers at Mimecast have reported a huge burst of a new type of fraud: sextortion scam caused by panic over security of smart cameras [12]. The scam has success because of the previously compromised home cameras including Google's Nest cameras, Amazon's Ring cameras and baby monitors, that among other studies shows that everything could be hacked [13].

Kaspersky Lab reported that more than 100 million attacks were targeted to IoT devices all over the world in H1 2019 [14] compared with 12 million in 2018. Taking advantage of the weak protection of IoT products, cybercriminals are putting more effort into creating and monetizing of IoT botnets. The report states that the attacks are not sophisticated and use exploits that allow botnets to compromise devices through old unpatched vulnerabilities and control them. Mirai and Nyadrop malware families remain the most popular techniques for compromising devices occupying about 40% of total each.

On September 2019 Trend Micro published the Uncovering IoT Threats in the Cybercrime Underground study [15] that describes how cybercriminal groups use IoT devices for their own purposes and what threats this poses. Trend Micro analysts investigated the darknet to find out which IoT vulnerabilities are most popular among cybercriminals. Hackers sell fresh vulnerabilities for routers, modified firmware for electricity meters. In the darknet they discuss hacking gas stations, sell and buy botnets based on IoT devices for organizing DDoS attacks. IoT devices as smart TVs, game set boxes are proposed to be used as cryptominers.

After all these attacks the companies are gradually realizing the consequences of not properly protected IoT devices and processes, but it is still not priority for business. The step forward the IoT security is accepting that they need support in understanding IoT technology that results into partnership with cloud service providers and IoT service providers. Companies most often cite the lack of competence as the main reason for such external partnership (47%), and then help and accelerate the deployment of the Internet of things (46%) as claimed in the Gemalto report about IoT security [16].

Although such partnerships can benefit businesses in implementing the IoT, organizations acknowledge that this is challenge for them because they do not have full control over the data collected by the IoT products or services when this data is transferred from partner to partner, which potentially leaves data unprotected.

Among other challenges we found there are lack of awareness in IoT security, lack of in-house expertise and support from the top management, undefined threat surface and undefined security level for IoT products and processes, while latter one is the consequences of lack of unified standards and legal regulatory documentation. More detailed the challenges are described in the next section.

## III. Challenges with IoT Security

The following list outlines the challenges that enterprises are faced during planning and deploying IoT security.

### A. *Non-Trusted Third Parties*

According to the report of Gemalto report about state of IoT security [16] most companies see the challenges with trying to secure their IoT products and services in ensuring data privacy and amount of data being collected. This user data can be shared between or even sold to various companies, violating the rights for privacy and security. Since data have a long way from its producer to the end consumer, including cloud providers, communication providers, IoT service providers, most of the companies consider third-party risk as a serious threat to sensitive and confidential information. This stated in the report of Ponemon about State of Cybersecurity in Small and Medium Sized Businesses [17] with numbers of 57 percent who consider that third parties expose their companies to risk regarding a data privacy and data breach, and 58 percent who are not confident that their primary third party would notify them if it had a data breach.

### *B. Lack of Awareness*

Regardless the fact that more than half (54%) of consumers own an IoT device (on average, two devices per person), only 14% consider themselves knowledgeable about the security of these devices [16]. This knowledge includes awareness about security measures, and principal understanding of what measure mitigate or eliminates what risk. Such statistics show that both consumers and enterprises need additional education in this area.

### *C. Unknown Threat Surface*

The biggest mistake of the businesses that data from IoT system is often not considered critical until it is used for billing and accounting. In their opinion, the device sending sensor measurement periodically does not carry critical information and is not of interest to hackers. Organizations state that it is hard to imagine what kind of threats towards IoT product could be, and they do not have many examples to learn from. The report [8] showed that number of companies that have no confidence in identifying assets for threat model [18,19], as well as in understanding and assessing cyber risks, raised from 9% in 2018 to 18% in 2019 which is caused by the emergence of new technologies like IoT, blockchain, big data, etc., that brought the complexity of an organization’s technology footprint, including threat and cyber risk assessment.

### *D. Lack of Support from Top Management*

Regarding the level of investment in security, the survey [16] showed that IoT device manufacturers and service providers spend only 11% of their total IoT budget on securing IoT devices. Regardless the 92% of companies have seen an increase in sales or use of the product following the implementation of IoT security measures, the company leaders are not encouraged by the widespread use of IoT security, they are more interested in getting their products to market quickly, rather than taking the necessary steps to build security in from the start.

Top managers pay attention to security in cases when IoT system is dealing with personal sensitive information, as customers’ medical, financial, or tax records, otherwise the security is out of priority of management. Security financial investment, especially advanced, is painful for top management, therefore, the level of security measures remains low in IoT and leads to so many successful attacks.

### *E. Lack of In-House Expertise*

Since IoT services is a new technology, for most of the companies is unknown territory that requires additional competence, and a finding an expert that skilled enough in IoT solutions can be challenging. For example, security professionals need practical knowledge of embedded devices, sensors, and computer-computer data communications, they should have experience in integrating heterogeneous protocols for data transfer, communications and network design both within the local Internet of things infrastructure and in cloud environments.

According to IoT Signals report from Microsoft [20] about half of the companies (47%) do not have enough workers skilled in IoT, and 44 percent are not having enough finances to invest in IoT security training for their employees. Lack of external guidance/regulation on how to secure IoT devices and services, lack of internal knowledge of how to provide security measures were pointed as challenging in dealing with IoT. Moreover, inside organizations it is not clear who is responsible for IoT security and who does what: responsibilities and competencies are fragmented within the companies that causes uncertainty among companies and customers.

### *F. Undefined Metrics for IoT Security*

The studies showed that companies really recognize the importance of protecting the IoT devices and data that they generate or transmit, and 50 percent of companies provide security based on a design approach. Two-thirds (67%) of organizations report using encryption as the main method of protecting IoT assets with 62% data encryption immediately upon reaching the IoT device, and 59% upon exiting the device [16]. But at the same time the organizations state that it is hard to define the right level of security, determine when that's fine enough. Basic encryption is good but this is an artifact measure inherent to IT security in general, however, more specific measures to ensure the security of IoT devices and processes are not popular due to a lack of understanding of the features of IoT systems, as limited memory and computational resources of IoT devices [21], special communication and information exchange protocols, supply chain complexity and increased connectivity of IoT ecosystems, as well as not understanding the very essence of these measures.

### *G. Lack of Standards*

The security of the Internet of things suffers from a lack of generally accepted standards. All businesses revealed the lack of standards, guidelines and/or checklists on how to ensure security of IoT [8, 16,17]. Adding new devices or their components to the IoT ecosystem given that there are no standards, increase the risk of penetrating into critical systems (e.g. industrial, municipal, energy, etc.) by intruders with the subsequent termination of operations.

Although, there are many best practices and recommendations for IoT security from the security-focused organizations, there is no single coherent structure. Large vendors, world leader companies have their own specific standards, while each IoT domain has its own incompatible standards from industry leaders in certain domain. The variety of these standards makes it difficult not only to protect systems, but also to ensure interoperability between them.

## IV. Measure to Strengthen IoT Security

The following list outlines the measures for the companies that reduce risks related to IoT security.

### A. Investment

Increasing investment into IoT security carries almost unlimited potential benefits in rise of protection, operational efficiency and in creating trustful relationships with customers. As survey [22] showed that the performing of better investment in the security allows for the business to stop more attacks, find and fix breaches faster and have less breach impact.

### B. IoT Security as Aart of Cybersecurity Business Strategy

The changing of business strategy forward new technologies trends related to digital transformation allows to achieve greater efficiency while also better protecting the business. However, in the process of including IoT development to the business strategy the organizations should not forget about the risks associated with IoT. Internet of things security, as part of cybersecurity policy, must be woven into corporate strategy, product design, budgets, and permeated with everyday business activities. Companies are required to change the approach to information security and the nature of their IT budgets, move their security mindset from technology-based defenses to new models for the implementation of information security, to proactive steps that include technology, process, and education.

### C. Outsourcing Security Operations to Third-Party

Most companies (99%) feel insufficient expertise to ensure the security of their products and processes, so they attract external consultants [23]. Using external suppliers and consultants in security operations can significantly increase the level of service and products without investment in technology or expert hiring.

An outsourcing continues to be popular solution in providing security measures for the companies: they prefer to outsource the security operations related to IoT, even if they have expertise, to do some operations as risk assessment, monitoring the traffic for malicious activities, incident response service. The outsourcing is more common trend among small and medium businesses, that was observed in previous years [24].

### D. Allocating Responsibility Within IoT Ecosystem

Nowadays, all businesses are not standalone production but complicated enterprise ecosystem with set of hardware, software and services. The potential breaches occurred in the company will affect not only company itself but hardware/software manufactures and all level of society. Cybersecurity could be one of these managed services that helps the company to tackle the IoT security risks. Third-party supplier can play a responsible role on helping the companies to protect against cyberattacks and providing security training for employees. Therefore, it is important to map the responsibility within all interacting elements in company's IoT ecosystem to specify and divide duties and responsibilities.

### E. Allocating Responsibility Within Company

Having cybersecurity team inside company with allocated task related to IoT security can improve cyber resilience, provide faster incident detection, shorter response time and in-time recovery process. Well organized, supported and managed by company leaders IoT security will help to deal with the pervasive risks of the IoT technology for business.

### F. Implementing of IoT Security Measures

After series of the attack and misusing of IoT devices the companies are forced to add security measures to their products or include into already running processes. The implementing of best practices and security measures as stated by ENISA in [25, 26] can help ensure overall security of IoT system and devices, prevent or properly respond to potential cyberattacks. There two approaches of implementing security measures to the product: at the design stage for new customers, and after the product is on the market. The first approach is the most effective and secure.

Both approaches can be accompanied by a systematic implementation or driven by customer requirements. During systematic implementation of IoT security the process is starting with threat modeling, risk assessment, and required security measures towards components of product and ending with mitigation, planning, and the optimal solution for each customer. But many companies admit that selection of IoT security measures is primary driven by customer requirements, and that some customers are not security-driven at all, they just need to have their data collected by IoT.

### G. Standardization and Legacy Regulation

The legal standards and regulatory frameworks aimed at IoT service providers and manufacturers, with large fines and working instructions, can raise responsibilities of the business for IoT security, as well as, both non-trusted third parties and not defined IoT security metrics challenges can be resolved with it. The set of dedicated compliance and standards how to handle and store sensitive IoT data can help with ensuring protection of user data and lead to more trust towards third parties who have access to the data.

Standardization and legacy regulation will be a driving factor in the development of cybersecurity hygiene and culture, raising awareness and responsibility. Mandatory set of measures and requirements for the security level in different IoT domains will increase customer confidence towards manufacturers of IoT products and services. Moreover, companies will no longer be unaware of what a sufficient level of security is, and there will be no need in search of an individual solution for each client that will allow save time and resources. The certification procedure for IoT devices should not be bureaucratic and provide the buyer with a guarantee that it has a certain degree of protection against hacker attacks. A certificate of quality can be issued both nationally and internationally, in the future. To begin with, the need for a security certificate can be indicated when conducting public and corporate procurements.

### H. Raising Awareness

Raising awareness about security of companies is one of the measures to improve IoT product security standards. Many authors and reports [9, 27] emphasize higher general awareness among customers and business can drive a market growth, increase the understanding of cybersecurity and data privacy. A high level of competences will create a more skilled workforce that can serve as a differentiator by itself.

## V. Analysis of IoT Security Measures

In this section the result of the analysis of measures for strengthening IoT security and risks associated with their implementation will be presented.

### A. Investment

A number of security reports from Ponemon, Accenture, Deloitte, Hiscox, PwC [17, 22, 23, 28, 29] have already noticed that in the past 5 years the companies have begun to pay more attention to security, have larger percentage of investments in security. The report of 2018 [28] showed the average spending of 10.5% of budget on cybersecurity programs, smaller companies with fewer than 250 employees spend 9.8% on average versus 12.2% for larger companies. The same trend of investment distribution showed another report of 2020 [30] with 10.9% in average and 11.2% for large companies that is insufficient in comparison with the potential impact of the cybercrime.

Although eighty-three percent of organisations agree that new technologies are necessary and crucial, investment is lagging. Only two out of five companies invest in new technologies, including IoT. However, companies are ready in the near future to increase investment in security of Internet of things: about half of the companies expressed a desire to do this, of which the most interested in investing were areas such as the automotive industry, industrial goods and technology [29].

### B. IoT Security as Part of Cybersecurity Business Strategy

Implementing IoT security as part of cybersecurity business strategy can help strengthen the security of IoT products and processes. But first, organizations need to change their approach to security because existing security strategies in the form of security appliance (FW, anti-virus solutions, intrusion detection systems) that perform technology-based defense functions are becoming not enough. All types of companies, including large businesses, continue to struggle with insufficient, outdated security strategies and plans that do not consider fully all relevant risks and threats. Standard approaches to information security focus on detecting threats and minimizing damage, rather than making digital products and processes secure by design. More than half of companies cites that the implementing of security measures is dictated by compliance with external regulations and policies, in 2017 only 55% of companies had security in their business strategy [31].

There is no research that can show business strategy of the companies towards the IoT security but mindset regarding common security strategy in the company that there is three way of focus: security operations operate under stealth and secrecy (60%), security efforts prioritize external threats (55%), security efforts mainly focus on prevention (55%) [31]. In total, 42 percent of companies have no governance policies associated with IoT risks included to the business strategy or to business continuity plan [22].

The most common reason why these enterprises do not consider it necessary to include security into the business plan is because they consider themselves too small or insignificant to justify such measures. The opinion that prevails in this category of respondents is that their IoT system will not be affected by cyberattacks.

The second popular reason for business is that cybersecurity is not considered enough in priority. Another cause is the unreasonable implementation of security from the point of view of financing when security measures increase the price of IoT product or process, which is not in favour of the management or the end user.

Quite many of companies do not consider security enough of priority and prefer to place functionality of the products and processes related to IoT on the higher level than security.

### C. Outsourcing Security Operations to Third-Party

The organizations believe that outsourcing is a cost-effective way to attract additional expert knowledge since it is quite difficult to convince management of in-house investments in such a narrow sector as IoT security.

The types of security services requested by companies from security suppliers can be divided into two types: outsourcing that oriented on providing certain service, and outsourcing that focused on the whole product. This means that companies do not have specific knowledge about the real products and tools that their external suppliers use. Some companies purchase just additional pentesting of the product in addition to the pentesting already conducted inside company, and some purchase all spectre of services during transmission, hosting and processing of data, including server security, databases, authentication and authorization of users for granting access to data collected by IoT devices.

The analysis of sources showed that the vast majority of companies (77%) have awareness of growing risks related to Internet of things, but only half of them (58%) are covered by effective security measures. The gap can be explained with lack of competence that varies from basic concept to matured experts who know and can implement security measures into operation process. Lack of competence and not understanding of how to protect their companies against cyberattacks leads companies to use help of external security providers, therefore, the outsourcing rate is so high. To strengthen IoT security competence the companies need more in-house expertise, hence, more budget should be allocated to this. But when it comes to investing top management needs a strong justification for driven factors why they need a security expert.

From the analysis of the reports we can conclude that the reasons for outsourcing are not only the lack of expertise in IoT security, but also the lack of time or human resources, therefore majority (93%) of companies indicated that they turn to suppliers in providing more than 10% of security operations, vulnerability management and incident monitoring. Some companies choose rather to completely outsource the area of security because suppliers are more efficient and often can deliver at a higher service level, some sought external help only for certain functions, such as setting up a firewall.

But outsourcing can introduce risk of trusting to third-parties: their failures could lead to some of cybersecurity breaches. Only 8% of companies are highly confident in external suppliers and 55% stated that they are fairly confident [8]. Therefore, with such a low level of trust, it is better to have internal expert also in-house with at least basic level of understanding the security measures, and proper evaluate what can be outsourced. According to Deloitte report [23], 16 percent of companies outsource more than 50% of their cybersecurity operations, and 65% outsource cyber operations in the 21–30% range.

Generally, the involvement of professional service providers or security consultants should be considered as positive aspect that gives confidence in ensuring cyber and IoT security, and helps to evaluate specific products.

### *D. Allocating Responsibility Within IoT Ecosystem*

The allocating IoT security operations to the external supplier demonstrates the trust relations inside value chain, and in many cases, relieves liability from the company itself. Another approach of managing security in the company is to do it with its own efforts and do not delegate security operation and trainings to outsourcing company. The allocating IoT security to the department or person in the company demonstrates a willingness to move towards including IoT security into the business strategy. In this case, all responsibility in providing protection measures and consequences of system breaking lays on the company itself.

The Gemalto report about state of IoT security shows that there is no clear understanding who is responsible for what operation in IoT system deployed in the company. If with responsibility for stored in the cloud IoT data all is clear, and cloud service provider is currently responsible for IoT security in the cloud, then the responsibility for other stages of operation of the IoT system are split between manufacturers of IoT products, IoT service providers, API developers and third-party security suppliers and specialists (e.g. Gemalto, Symantec, Verizon etc.)

The splitting of responsibilities and concrete allocation between suppliers will allow to handle proper the incidents with information leak or in the event of a serious attack, when the likelihood of litigation is high, and it is the leaders of the company who will have to prove to the stakeholders and customers that they did everything they could to protect the business, data, and product and explain why it was not possible to effectively prevent the incident.

### *E. Allocating Responsibility Within Company*

For a long time, cybersecurity has been the responsibility of IT departments. The most common misconception among business leaders is that they believe that Information Security is part of IT. But security is a separate area that requires time.

In many cases the task of implementing, monitoring and testing the IoT security of products and processes is done by the person who just interested in this field. This person has no expertise in security or has little, but spend his/her time on getting knowledge on the specific topic and implementing security measures. This situation usually occurs when current employees do not have the right expertise to carry out IoT security as required. This approach is more appropriate to medium and small companies than to large business mostly due to lack of the resources.

Regardless the high concern about cybersecurity of IoT (80% think that a security incident related to unsecured IoT product could be catastrophic [17]), the top management rarely participates in cybersecurity discussions regarding, for example, building security into product designs. Earlier research showed that only 22% of companies have business leaders are accountable for cybersecurity [22], and only 21% monitor the risk of their IoT products [17]. But even these numbers need to be shifted more towards responsibility of top management because nowadays the cybersecurity is becoming a common task for all company employees. Top management must allocate the necessary resources and participate in the development of proactive and integrated response programs for cyberattacks.

### *F. Implementing of IoT Security Measures*

Regardless the most companies (80%) are interested in IoT security [17], they are not in a rush with implementing security measures. According to [8] almost one-quarter of companies is aware of cybersecurity risks, but some companies do not have IoT security measures in place at all. This is because with the adopting of new technologies (IoT, AI, block chain, cloud computing), the main preference for half of companies (50%) was the pushing ahead a digital transformation, despite the potential security risks associated with them [8].

Since IoT is a new trend, after series of the attack and misusing of IoT devices, the companies are forced to add security measures already after a product is produced or include into already running process. Some companies have hung with a basic security measures as encryption and passwords and do not progress more due to the lack of awareness and guidance of how to do it. For example, every second company (52%) implements their security measures based on the current cybersecurity needs, and do not consider future pervasive risks and needs [22]. Some companies have more than basic level with a systematic approach based on one of the cyber security frameworks, e.g. ISO 27001 [32], IoT Security Foundation [33], NIS directives [34], NIST SP 800-53 [35], IEC 62443 [36] UL 2900 series [37], and Cyber Essentials in UK [38].

The most important security measures in IoT should be focused on authentication, secure communication, secure handling, secure storage, which are aimed to identification of the communicating devices, protecting data in transit, protecting data in process, and protecting data at rest, respectively [39]. All companies should strive for providing all set of the data-centric security measures. This is not only correct, but also important if organizations have serious intension towards protecting their assets and the data of customers [18, 19, 25].

Basic security measures have to include the form of encryption and password policy. But it is required to keep balance between cryptography algorithm and level of security. Some of the algorithms are high energy consuming that can reduce the lifetime of IoT devices that powered by battery [40].

Companies are more focused on their customers and driven by them in choosing the security measures. Customers stimulate demand for IoT security in the products or services that companies provide. The more aware customers about security risks and threats, the bigger the demand for IoT security measures. Therefore, it is very important to help consumers really understand what is happening with their data and teach them how it is possible to protect it.

### *G. Standardization and Legacy Regulation*

While certification promises a solution to many problems for business and for consumers of IoT services, a unified standard still does not exist. However, the analysis of attacks on IoT devices and the most frequent vulnerabilities that led to a successful attack forced the organizations focused on providing guides for security to make a number of recommendations on protecting IoT devices and IoT infrastructures.

Among bodies involved in the producing of recommendation for security of IoT devices, such as OWASP, ENISA, IoT Foundation, NCSC [33-35]. The recommendations partially duplicate each other, are advisory in nature, therefore, cannot be considered as standards that must be followed, and most companies simply ignore them or do not consider as legacy regulation.

Due to the fact that passwords are the most common weakness, the principal recommendation in all guidelines relates to the strengthening and control of the use and procedure of generating passwords. The security problems stemming from default or weak passwords on IoT devices are well known and were discussed in the Section 2.

To enhance security, all passwords of user IoT devices should be unique and without the ability to reset them to "universal" factory settings. Manufacturers are required to provide a public point of contact so that everyone can report vulnerabilities and count on "timely action." This measure is not new in the field of Internet products: the IETF provides draft [41] that explains using of the file security.txt with contacts for reporting the security vulnerabilities. It is still recommendation and draft regulations but it is way to interact customers with manufactures or service providers. This measure will reduce the vulnerability in other devices of the same series or versions due to timely replacement or patching of the device. A vulnerability identified in time will bring closer the moment of taking protective measures for product security.

Another measure of enhancing cyber security of IoT stated in the many guidelines is the ability to provide product updates, either remotely or in place. The using of old versions of software are identified as high security risk factors in many regulations from organizations aimed at ensuring IoT security. Lack of secure updating mechanism is ranked fourth in the OWASP top 10 IoT security classification.

Given the heterogeneity of applications by end-user (private person, company, state) and application domain (health, automotive, HVAC), the system of standards will be multi-level, varying in degree of coverage and detail [42].

While mitigating the risk associated with the low trust to suppliers the appearing of new standards and regulation can pose another challenge related to compliance with various standards, legal and regulatory structures.

### *H. Raising Awareness*

Managing IoT dynamic risks will be a challenging task for organizations whose employees lack the qualifications in creating, implementing, and troubleshooting an IoT security system. Therefore, the IoT security training for employees, customers and business leaders is required to build effective cybersecurity culture. The results of such a security training will be most effective when the issue of the importance of information protection of the business environment is put as a priority by executive management and initiative is coming from the organization's leadership.

More skilled workforce will provide a high level of competence and awareness and will be a plus for the company, as well as create market attraction, as consumers will demand more cybersecurity in the products they buy.

IoT security training has not yet been widely disseminated in most organizations. The percentage of organizations that conduct educational training for employees and third parties about the risks created by IoT devices remains small - only 24% of companies currently provide such information and education [17]. However, it is promising that three out of ten business leaders plan to invest more in IoT security training in the future [29].

Finally, the effectiveness of each security measure and its impact towards the identified challenge is presented in Table 1.

TABLE I
IMPACT OF THE MEASURES ON THE ISSUES WITH IOT SECURITY

| | Third parties | Lack of awareness | Threat surface | No support from management | No in-house expertise | Undefined metrics | Missing standard |
|---|---|---|---|---|---|---|---|
| Investment | Medium | High | | Medium | High | Low | |
| IoT security strategy | | Medium | High | High | | Medium | Low |
| Outsourcing | | | Medium | | | | |
| Ecosystem leadership | High | | | High | Low | | |
| Company leadership | Medium | | | High | Medium | | |
| IoT security measures | | | High | | | Low | Medium |
| Standards | | | | | | High | High |
| Raising awareness | | High | High | | High | | Medium |

Regardless its popularity the outsourcing is not effective strategy of the company in solving the issues related to IoT security. This measure leaves the company blind in relation to the threat surface, possible security solutions, does not increase knowledge within the company.

The most effective method for solving a series of the most cutting-edge problems is an investment but not all companies are willing to spend more on security and on staff education.

Raising awareness is one more effective way to solve the set of the issues related to personal expertise inside company. During training the employees can expand their knowledge about IoT technology, attack surface, protection security measures, and, also, help the client to select IoT solution.

Using this matrix, the companies will be able to navigate in the selection of security measures and choose the most effective way to solve their specific issues.

## VI. Conclusion

The paper was aimed at identifying challenges with IoT security for business and finding possible measures to overcome these challenges. The proposed mapping of the impact of each of the security measure will help companies change their mindset towards IoT security, increase the protection of devices, processes and customers data, and thus, business competitiveness.

Our findings continue to highlight the importance of implementing the IoT security as part of the business strategy. Among the reasons why companies have difficulties with creating a stronger IoT security posture are 1) lack of in-house expertise, 2) not understanding how to protect against IoT cyberattacks, 3) not a priority issue, 4) lack of collaboration with other functions, 5) management does not see cyberattacks on IoT products as a significant risk. First two are primary reasons for not implementing the IoT security, others are main reasons for companies to consider IoT security as afterwards because operational processes considered as more important.

Only two of ten companies monitor the risk of their IoT products and processes. This is catastrophic because these products go to the market, start operate in the customer houses or monitor engines without any cyber security check. To correspond to the future cyber resilience needs the companies must update cybersecurity strategies and include IoT security.

Implementing IoT security into the business strategy will help raise the level of strategic awareness on all levels, as well as provide companies with the commercial, organizational and technical qualifications for strategic and commercial use of IoT security considering it as a competitive parameter. Standardization and regulatory control of IoT security will make the security of the Internet of Things tangible and understandable for business and, therefore, uses IoT and IoT security as a driving force for growth.

Most organizations have a perception of IoT security, sources of cyber risk, mechanisms of ensuring security, but they feel insufficient confidence in order to provide value chain related to IoT security themselves. Moreover, it seems that companies often expect the information about IoT security to be propagated to them from security profiling or security governance organization, rather than to seek it out themselves. Due to lack of awareness and confidence, companies use external suppliers and consultants in IoT security operations. The organizations believe that it would be useful to get more recommendations or checklists to provide IoT security.

One of the challenges is related to implementing security measures, they are either basic in form of encryption and secure data transmission or have necessary for customer level. Such a low level of security measures is due to the fact that more advanced security mechanisms are required more financial investment that is painful for top management. The organizations require more accurate estimation of costs associated with a cyberattack on IoT products and processes, hence providing greater clarity around the potential financial risks.

Finally, while more organizations are concerned about IoT security and consider it critical for business and necessary for business development at age of growing digitalization and within trend of pervasive connected devices, the analysis showed that implementation of IoT security is not a priority task. Increasing understanding of IoT security risks and threats among top management and business leaders may facilitate the change of strategy towards inclusion of IoT security in the company's priority tasks. The companies must update the way they plan and execute IoT security.